\documentclass[11pt]{article}

\usepackage[margin=1in]{geometry} 

\usepackage{setspace} 
\usepackage{rotating} 
\usepackage{amsmath} 
\usepackage{amsfonts, amssymb} 
\usepackage{graphicx} 
\usepackage{epstopdf} 
\usepackage{hyperref} 
\usepackage{bm} 
\usepackage{bbm}
\usepackage{multirow} 
\usepackage{latexsym} 
\usepackage{natbib}
\usepackage{rotating} 
\usepackage{titlesec} 
\usepackage{caption} 
\usepackage{mathtools} 
\usepackage{color} 
\usepackage{enumitem} 
\graphicspath{ {plot/} }
\usepackage{diagbox}
\usepackage{pict2e}
\usepackage{etoolbox}
\usepackage[english]{babel}

\newcommand{\reals}{\mathbb{R}}

\usepackage[ruled,vlined]{algorithm2e}
\include{pythonlisting}
\usepackage{multirow}

\newcommand{\bX}{\mathbf{X}}

\newcommand{\bz}{\mathbf{z}}

\newcommand{\cond}{\, \vert \, }

\begin{document}

\title{Propensity Score Modeling: Key Challenges When Moving Beyond the No-Interference Assumption}

\author{
  Hyunseung Kang$^1$, Chan Park$^2$, and Ralph Trane$^1$
  \\[0.5cm]
  {\footnotesize $^1$: Department of Statistics, University of Wisconsin--Madison}\\
  {\footnotesize $^2$: Department of Statistics and Data Science, The Wharton School of Business, University of Pennsylvania}
  \\[0.5cm]
  {\small All authors are ordered alphabetically and all authors contributed equally to this work} \\[0.5cm]
}

\date{}


\maketitle

\begin{abstract}
The paper presents some models for the propensity score. Considerable attention is given to a recently popular, but relatively under-explored setting in causal inference where the no-interference assumption does not hold. We lay out some key challenges in propensity score modeling under interference and present a few promising models based on existing works on mixed effects models.
\end{abstract}


\section{Introduction}
Ever since \citet{RR1983} (referred to as RR), propensity scores and, more generally, balancing scores have played a central role (no pun intended) in not only observational studies, but also every aspect of causal inference. Some of the most celebrated properties of the propensity score include:
\begin{enumerate}
\item the propensity score balances covariates between the treatment and the control group 
(i.e., Theorem 1 of RR);
\item the propensity score is the coarsest balancing score whereby any balancing score can be transformed into the propensity score (i.e., Theorem 2 of RR)
\item if the treatment is strongly ignorable given pre-treatment covariates, the treatment is strongly ignorable given the propensity score (i.e., Theorem 3 of RR).
\end{enumerate}
In addition to these ``large sample'' properties where the propensity score is known, RR and several others have explored various  ``small sample'' properties where the propensity score must be modeled and estimated; see Section \ref{sec:ps} for a brief review. However, there are far fewer works on the propensity score (or a related object) when one of the core assumptions underlying RR, the no-interference assumption, does not hold. The theme of the paper is to discuss this growing, but surprisingly under-developed area on propensity score modeling under interference.

Briefly, interference  \citep{CoxInterference1958} is a phenomenon where the response from one study unit may be affected by the treatment status of other study units. For example, when evaluating vaccination campaigns, an individual's vaccination status (i.e., treatment status) may not only affect their own response, say disease status, but also their peers' responses; see \citet{HHReview2016} and references within for more discussions. 
RR and many other works that have followed often assumed the no-interference assumption via the Stable Unit Treatment Value Assumption \citep{SUTVA1980,SUTVA1986}. But, during the past 20 years, there has been an explosion of works on relaxing the no-interference assumption (e.g., \citet{hong2006evaluating,Rosenbaum2007Interference,HH2008,TTVanderweele2012,van2014causal,ogburn2014causal,choi2017estimation,Liu2016,Liu2019,basse2019,savje2021average,forastiere2021identification,Park2022_EffNet}). Notably, some (but not all) of the ``large sample'' properties of the propensity score (or an equivalent object) have been established under various types of interference, especially partial interference \citep{Sobel2006} where interference occurs within pre-defined, non-overlapping clusters of study units \citep{HH2008, TTVanderweele2012,Liu2016,forastiere2021identification,Park2022_EffNet}. However, in our opinion, there are few (if any) works on the ``small sample'' properties where the propensity score is not known a priori and must be modeled.

The rest of the paper discusses key challenges in modeling the propensity score (or an equivalent object) under interference. We start by discussing the case where the no-interference assumption holds (i.e., RR's original setup). Specifically, we discuss the ``go-to model'' for the propensity score (e.g., logistic regression) and some recent extensions using machine learning and cross fitting \citep{DML}. Subsequent sections discuss the case where the no-interference assumption does not hold, with Section \ref{sec:ps_pint} exploring the case under partial interference and Section \ref{sec:ps_gint} exploring the case under general interference (i.e., interference that is not partial interference). In both cases, we stress how under interference, each study unit's observed data is not i.i.d. and the statistical dependency in the observed data could be too unwieldy without some strong modeling assumptions. More broadly, our goal in these sections on interference is not to provide a one-size-fits-all or a go-to model for the propensity score, akin to the case under the no-interference assumption. Rather, we want to bring to light some key  challenges that investigators in causal inference should be aware of when studying the `small sample'' properties of the propensity score under interference.

\section{A Review of Propensity Score Modeling Without Interference}
\label{sec:ps}
When interference is absent, the most common modeling assumptions for the propensity score are that (a) each study unit $i$'s binary treatment $Z_i \in \{0,1\}$ and $p$ pre-treatment covariates $\mathbf{X}_i \in \reals^p$ are i.i.d. and (b) the propensity score follows a logistic regression model. Formally, let $e(\mathbf{X}_i) := P(Z_i = 1 \mid \mathbf{X}_i)$ denote the propensity score.  We can restate the two assumptions (a) and (b) as:
\begin{align}
P(Z_1\ldots,Z_n \mid \mathbf{X}_1,\ldots, \mathbf{X}_n) &= \prod_{i=1}^{n} P(Z_i \mid \mathbf{X}_i), \quad{} Z_i \in \{0,1\}, \mathbf{X}_i \in \reals^p, \label{eq:iid} \\
e(\mathbf{X}_i):= P(Z_i=1\mid \mathbf{X}_i) &=  \big\{1+ \exp(-\mathbf{X}_i^\intercal \bm{\beta}) \big\}^{-1}, \quad{} \bm{\beta} \in \reals^{p}. \label{eq:logit} 
\end{align}
Equation \eqref{eq:iid} corresponds to the i.i.d. assumption and equation \eqref{eq:logit} corresponds to the logistic regression model assumption. The parameter $\bm{\beta}$ in model \eqref{eq:logit} is often estimated using the maximum likelihood principle and the estimated propensity score, denoted as $\widehat{e}(\mathbf{X}_i)$, replaces the parameter $\bm{\beta}$ in equation \eqref{eq:logit} with its maximum likelihood estimate. 

Some works have replaced the logistic regression model with less parametric models, such as tree-based models (e.g., \cite{mccaffrey2004propensity} and \citet{lee2010improving}) or neural networks (e.g., \citet{Shi2019} and \citet{Farrell2021}); see Table 1 of \citet{westreich2010propensity} for a summary. Formally, these works still maintain equation \eqref{eq:iid}, but replace equation \eqref{eq:logit} with an assumption that the true propensity score $e$ belongs to some ``well-behaved'' family of functions, denoted as $\mathcal{F} = \{f(\mathbf{x}): \reals^p \to [0,1]\}$; one of the most popular ``well-behaved'' family of functions in propensity score modeling is the Donsker class\footnote{Donsker family of functions include all bounded functions, monotone functions, and many parametric functions (including logistic regression models), to name a few; see Chapter 19.2 of \citet{van2000asymptotic} for additional examples.}. Additionally, these works assume that the squared (i.e., $\ell_2$) distance between the estimated propensity score $\widehat{e}$ and the true propensity score $e$ goes to zero at a certain rate as the sample size $n$ increases, i.e.,
\begin{align} \label{eq:ml}
 \|\widehat{e} - e\|_2^2 :=  \int \left\{\widehat{e}(\mathbf{x}) - e(\mathbf{x}) \right\}^2 P(\mathbf{X}_i = \mathbf{x}) d\mathbf{x} = O_p(r_n^2) \text{ where } r_n \to 0 \ .
\end{align}
The term $r_n$, which is a positive, non-random sequence of numbers, measures how fast the $\ell_2$ distance between the estimated $\widehat{e}$ and the true $e$ goes to zero as the sample size $n$ increases.
For example, $r_n = n^{-1/2}$ indicates that as $n$ increases, the distance between $\widehat{e}$ and $e$ (i.e., $\|\widehat{e} - e\|_2^2$) is approaching zero at a rate that is comparable to how $n^{-1/2}$ is approaching zero. The notation $O_p(r_n^2)$ is a more mathematically concise statement of this explanation where $O_p(r_n^2)$ states that the ratio $\|\widehat{e} - e\|_2^2/r_n^2$ must be bounded in probability\footnote{For readers who are not familiar with the idea of measuring distances between two functions, roughly speaking, the expression $\|\widehat{e} - e\|_2^2$ can be interpreted as computing the squared distance between the estimated propensity score $\widehat{e}$ and the true propensity score $e$ at each point $\mathbf{x} \in \reals^p$  and taking a weighted average of these distances across $\mathbf{x}$ where the weight is the probability density of $P(\mathbf{X}_i = \mathbf{x})$. Also, unlike more familiar notions of distance between two points where the distance between point A and B are deterministic and fixed, the distance $\|\widehat{e} - e\|_2^2$ is a random quantity, specifically a scalar random variable, because the estimated propensity score $\widehat{e}$ depends on observed samples, which are random realizations from a population. In other words, if we obtained another sample of data, we would end up with a slightly different estimate of the propensity score $\widehat{e}$ and a slightly different distance $\|\widehat{e} - e\|_2^2$. Finally, it's common in the literature to find a decreasing sequence of $r_n$ that would make the (random) ratio $\|\widehat{e} - e\|_2^2/r_n^2$ be bounded in probability; see Chapter 1.4 of \citet{lehmann1999elements} for details.}. Also, similar to a marathon where a fast runner is ``preferred'' over a slow runner, it is desirable to have $\|\widehat{e} - e\|_2^2/r_n^2$ go to zero (i.e., the finish line) as quickly as possible. Or, to put it in terms of $r_n$ which measures the rate at which the distance $\|\widehat{e} - e\|_2^2$ goes to zero, it is desirable to have $r_n = n^{-1/2}$ (i.e., a ``fast'' rate of convergence) rather than $r_n=n^{-1/5}$ (i.e., a ``slow'' rate of convergence).

Finally, in recent years, the assumption that the true propensity score belongs to some well-behaved function class (e.g., $e \in \mathcal{F}$ where $\mathcal{F}$ is Donsker) has been replaced with a slightly more restricted version of equation \eqref{eq:ml} where the estimator of the propensity score $\widehat{e}$ is restricted to those that use a version of sample splitting known as cross fitting \citep{DML}. Specifically, the first sub-sample of the data is used to learn the functional form of $e$ while the second, independent sub-sample of the data is used to evaluate $e$ (i.e., obtain estimated probabilities) for each data point in the second sample. Then, the roles of the two sub-samples are reversed and the final estimator of the treatment effect combines the evaluations of $e$ in each sub-sample; see \citet{DML} for details. As long as $\widehat{e}$ constructed using cross fitting converges to the true $e$ at a fast rate, the true propensity score $e$ can be in an ill-behaved function class $\mathcal{F}$\footnote{One famous example of a family of ill-behaved functions from \citet{DML} is a high dimensional linear regression model where the dimension of the covariates is growing with sample size $n$.}.  In other words, with cross fitting, investigators have to worry less about the well-behavedness of a function class. 

\section{Propensity Score Modeling Under Partial Interference: Some Progress} \label{sec:ps_pint}
\subsection{A Brief Review of Large Sample Properties}
As mentioned above, partial interference is a type of interference where study units are partitioned into non-overlapping clusters and interference only occurs within clusters. To describe the clustering structure intrinsic under partial interference, suppose we add one additional subscript $j$ to the variables introduced above so that $Z_{ij}$ and $\mathbf{X}_{ij}$ are the treatment and pre-treatment covariates, respectively, of study unit $j=1,\ldots,n_i$ in cluster $i=1,\ldots,I$; there are $I$ clusters and $n = \sum_{i=1}^{I} n_i$ study units in the study. Additionally, for each study unit $j$ in cluster $i$, let $R_{ij}(z_{ij}, \mathbf{z}_{i(-j)})$ be their potential outcome if their treatment is set to $z_{ij} \in \{0,1\}$ and their peers' treatments are set to $\mathbf{z}_{i(-j)} \in \{0,1\}^{n-1}$ where the subscript $(-j)$ indicates the vector $\mathbf{z}_i$ with the $j$th element removed. Note that the notation $R_{ij}(z_{ij}, \mathbf{z}_{i(-j)})$ encodes the partial interference assumption where the potential outcome of unit $j$ in cluster $i$ only depends on the treatment assignment of  units in cluster $i$, i.e., $z_{ij}$ and $\mathbf{z}_{i(-j)}$. Also, under the no-interference assumption, $R_{ij}(z_{ij}, \mathbf{z}_{i(-j)})$ collapses to $R_{ij}(z_{ij})$ where unit $j$'s potential outcome only depends on unit $j$'s own treatment assignment $z_{ij}$. 


Without further assumptions on the functional form of $R_{ij}(z_{ij}, \mathbf{z}_{i(-j)})$, 
the ``partial interference'' equivalent of the propensity score in RR is the cluster-level propensity score (CPS), defined as
\begin{equation}
e(\mathbf{z}_i \mid \mathbf{X}_i ) := P(\mathbf{Z}_{i} = \mathbf{z}_i \mid \mathbf{X}_i), \quad{} \mathbf{z}_i \in \{0,1\}^{n_i}, \ \mathbf{X}_i = (\mathbf{X}_{i1},\ldots,\mathbf{X}_{in_i}), \ \mathbf{X}_{ij} \in \reals^p. \label{eq:CPS}
\end{equation}
Under the ``large sample'' setting where the CPS is known, 
Section 5 of  \citet{TTVanderweele2012} and Section 3 of \citet{Liu2016} showed that various causal estimands under partial interference, such as the average direct effect and the average spillover effect, can be unbiasedly estimated. Critically, unlike the case where the no-interference assumption holds, the large sample properties under partial interference require knowing the entire probability distribution of $\mathbf{z}_i$ for each cluster $i$; simply knowing the expectation $E ( \mathbf{Z}_i \mid \mathbf{X}_i ) $ is not sufficient\footnote{Interestingly, this phenomena has also been observed in the setting where the treatment is continuous, but the no-interference assumption holds; see page 331 of \citet{joffe1999invited}.}. 

\subsection{Parametric Model: Mixed Effects Logistic Regression}
Modeling the CPS, which is a multivariate conditional probability distribution of $\mathbf{Z}_i$ given $\mathbf{X}_i$, is a far more difficult task than modeling the propensity score $e(\mathbf{X}_i)$ under the no-interference assumption. The list below lays out some key challenges that we (and others) have faced when modeling the CPS.
\begin{itemize}
	\item[1.] The treatment vector $\bz_i$ in the CPS can take on any value in the set of binary vectors of length $n_i$ and in real data,  especially when $n_i$ is moderate to large, it's unlikely that every permutation of the treatment vector is observed. 
	\item[2.] The CPS depends on the size of each cluster. For example, if cluster $i=1$ has two study units and cluster $i=2$ has four study units, the dimension of $\mathbf{z}_i$ in the CPS differs between the two clusters.
	\item[3.] For each cluster $i$, the joint distribution of $\mathbf{Z}_i$ is likely dependent. Relatedly, the ordering of study units in a cluster may be relevant.
	\end{itemize} 
To elaborate on the last point concerning ordering, suppose all clusters in the study consist of identical twins and only one of the twins is treated. If ordering matters, the probability that the first twin $j=1$ in cluster $i$ is treated may be higher (or lower) than the probability that the second twin $j=2$ in cluster $i$ is treated. In contrast, if ordering doesn't matter, say the two twins are exchangeable within each cluster\footnote{For each cluster $i$ of size $n_i =2$, we say that the conditional distribution of $P(Z_{i1},Z_{i2} \mid \mathbf{X}_{i1}, \mathbf{X}_{i2})$ is exchangeable if the distribution is identical after relabeling of the $j$th indices, i.e.,  $P(Z_{i1} = z_1,Z_{i2} = z_2 \mid \mathbf{X}_{i1}=\mathbf{x}_1, \mathbf{X}_{i2} = \mathbf{x}_2) = P(Z_{i2} = z_1,Z_{i1} =z_2 \mid \mathbf{X}_{i2} = \mathbf{x}_1, \mathbf{X}_{i1} = \mathbf{x}_2)$.  
},
 then the probability of the first twin being treated is the same as the probability of the second twin being treated.

The most popular solution  to address all of the above challenges is using a mixed effects logistic regression model \citep{TTVanderweele2012,PH2014,Liu2016,Liu2019,Barkley2020,Park2022_EffNet}, i.e.,
\begin{align}			\label{eq:glmm0}
& P(\mathbf{Z}_{1},\ldots,\mathbf{Z}_{I}  \mid \mathbf{X}_{1},\ldots, \mathbf{X}_{I}) = \prod_{i=1}^{I} P( \mathbf{Z}_i \mid \mathbf{X}_i ), \\
& P(\mathbf{Z}_i = \mathbf{z}_i \mid \mathbf{X}_i) 
= \int_{-\infty}^\infty \prod_{j=1}^{n_i} P(Z_{ij} = z_{ij} \mid \mathbf{X}_{ij}, V_i = v) P(V_i= v) dv, \ \mathbf{X}_{ij} \perp V_i, \ V_i \stackrel{i.i.d.}{\sim} N(0,\sigma_V^2), \label{eq:glmm1} \\
&P(Z_{ij} = 1 \mid \bX_{ij}, V_i) = 
\big\{ 1+\exp(-\mathbf{X}_{ij}^\intercal \bm{\beta} - V_i) \big\}^{-1}
. \label{eq:glmm2}
\end{align}
Equation \eqref{eq:glmm0} states that the clusters are independent from each other. Equations \eqref{eq:glmm1} and \eqref{eq:glmm2} formalize the mixed effects logistic regression model where the term $V_i$ is the Normal random effects term with mean $0$ and variance $\sigma_V^2$. The random effects term is critical as it effectively addresses all of the above challenges by (a) accounting for the correlation in the treatment variables among study units in the same cluster, (b) decomposing the joint distribution of $\mathbf{Z}_i$ into a parametric model that does not depend on the cluster size $n_i$, and (c) pooling information across different study units to better estimate some treatment probability vectors $\mathbf{z}_i$ that have not been observed in the data. We remark that the model parameters, specifically $\bm{\beta}$ and $\sigma_V^2$, are typically estimated with a maximum likelihood estimator or a restricted maximum likelihood estimator \citep{LME4}. 

\subsection{A Semiparametric Model}
Similar to how investigators went beyond the logistic regression model for the propensity score under the no-interference assumption, we can use less parametric variants of the mixed effects logistic regression model to reduce issues arising from model mis-specification. This section discusses one extension of the mixed effects logistic regression model based on works by \citet{NPGLMM1}, \citet{NPGLMM2}, and \citet{NPGLMM3}. More generally, to the best of our knowledge, these types of semiparametric models have not been used in practice and we hope some of the ideas discussed here could be useful for future investigators. 

Formally, suppose we maintain the assumptions in
\eqref{eq:glmm0} and \eqref{eq:glmm1}. But, instead of the mixed effects logistic regression model in  \eqref{eq:glmm2}, we adopt the following semiparametric model for $Z_{ij}$ given $\bX_{ij}$ and $V_i$:
\begin{align}		\label{Chan-EtaPS}
	P(Z_{ij} = 1 \mid \bX_{ij}, V_i) = \big[ 1+\exp \big\{ - f(\mathbf{X}_{ij}) - V_i \big\} \big]^{-1} , \quad{} f \in \mathcal{F} = \{f(\mathbf{x}) \mid \reals^p \to \reals\}.
\end{align}
Here, $f$ is a nonparametric summary of the covariates $\mathbf{X}_{ij}$ and belongs to some family of functions $\mathcal{F}$. The aforementioned works have assumed that $\mathcal{F}$ is a ``well-behaved'' family of functions. 
However, using ideas in Section \ref{sec:ps}, we can use cross fitting to estimate $f$ without having to make assumptions about $\mathcal{F}$ being well-behaved. Specifically, we propose the following procedure to estimate the CPS based on the semiparametric model \eqref{Chan-EtaPS}.
\begin{enumerate}
\item Estimate the ``original'' propensity score $e(\mathbf{X}_{ij}) = P(Z_{ij} = 1 \cond \bX_{ij})$ by using cross fitting in Section \ref{sec:ps}. Note that we would use
cluster-level cross fitting where entire clusters (not study units within clusters) are partitioned into sub-samples.
\item Estimate $f$ and $\sigma_V^2$ in \eqref{Chan-EtaPS} by iteratively solving a (numerical) integral equation for $f$ and the usual restricted maximum likelihood estimator for $\sigma_V^2$. Specifically,
\begin{enumerate}
\item By Bayes rule, $f$ and $e$ must satisfy the following integral restriction:
\begin{align}							\label{Chan-IntegralEquation}
e(\bX_{ij}) 
	=
	\int_{-\infty}^\infty \big[ 1+ \exp\big\{ - f (\bX_{ij})- v \big\} \big]^{-1} \phi(v,0,\sigma_V^2) dv
\end{align}
Here, $\phi(v,0,\sigma_V^2)$ is the density function of a Normal distribution with mean zero and variance $\sigma_V^2$. Given $\sigma_V^2$ and estimated $e(\mathbf{X}_{ij})$ from step 1, we can use any number of numerical integration techniques to find $f(\mathbf{X}_{ij})$ that satisfies \eqref{Chan-IntegralEquation}.
\item Given $f(\mathbf{X}_{ij})$, estimate $\sigma_{V}^2$ via the usual restricted maximum likelihood estimator from mixed effects logistic regression where $f(\mathbf{X}_{ij})$ is treated as a scalar covariate.
\item Iterate steps 2a and 2b until convergence. 
\end{enumerate}
\item Plug in the estimated values of $f(\mathbf{X}_{ij})$ and $\sigma_V^2$ into equation \eqref{eq:glmm1} to obtain an estimator of the CPS.
\end{enumerate}
We remark that there are works that allow for nonparametric distribution of the random effect term $V_i$ at the expense of having a parametric summary of the covariates, i.e., the opposite of the above semiparametric model where $V_i$ is parametric, but there is a nonparametric summary of the covariates. Unfortunately, these models are generally more complex to fit, suffer from efficiency loss, and require very strong conditions for identifiability  \citep{Agresti2004, GarciaMa2016}. Personally, in our own (admittedly limited) experience, we have found that a parametric distribution for the random effect is usually sufficient to capture the correlation structure of binary treated variables unless the correlation is extremely complex or heterogeneous across clusters. 
Instead, we find that the most problematic aspect of the mixed effects logistic regression model comes from the linear term $\mathbf{X}_{ij}^\intercal \bm{\beta}$ and the proposed semiparametric model above is a reflection of our (again, limited) experience with real data.

\section{Propensity Score Modeling Under General Interference: Many Challenges} \label{sec:ps_gint}
Broadly speaking, propensity score modeling under general interference is even more difficult than that under partial interference. The most challenging issue under general interference is that the propensity score (or an equivalent object) strongly depends on the sample size $n$ and consequently, any model is naturally high dimensional where every additional unit to the study can potentially add more parameters to the propensity score model. Also, like before, the data is not i.i.d. For completeness, we mention one existing attempt to address this challenge. However, unlike the setting under partial interference, there is a \emph{lot} of room for improvement in this area.

Before we begin, suppose we revert back to the notation in Section \ref{sec:ps} where we only use one subscript. Let $R_{i}(z_{i}, \mathbf{z}_{(-i)})$ be the potential outcome of unit $i$ given their own treatment $z_i \in \{0,1\}$ and their peers' treatments $\mathbf{z}_{(-i)} \in \{0,1\}^{n-1}$. Note that under no interference, the potential outcome $R_{i}(z_i, \mathbf{z}_{(-i)})$ collapses to $R_{i}(z_i)$.

 \citet{aronow_exposuremapping_2017} expanded on the ideas from \citet{manski2013identification} and introduced an exposure mapping to help with defining causal estimands and estimation of said estimands under general interference. The main goal behind an exposure mapping is to reduce the dimension of the peers' treatment status in the potential outcome so that identification and estimation of causal effects is tractable. A bit more formally, suppose we assume that the potential outcome of study unit $i$, $R_i(z_i, \mathbf{z}_{(-i)})$, only depends on the treatment assignment of unit $i$, $z_i$, and the value $g_i = g(\mathbf{z}_{(-i)})$, where $g: \{0,1\}^{n-1} \mapsto \reals$ is a fixed, known, correctly specified function that summarizes the peers' treatment assignments into a scalar quantity. For example, if the potential outcome of unit $i$ depends on unit $i$'s treatment assignment and the the total number of peers treated\footnote{This type of exposure mapping is so common in the literature that it has its own name, stratified interference \citep{HH2008}}, $g$ would be $g(\mathbf{z}_{(-i)}) = \sum_{i' \neq i} z_{i'}$ and $g_i$ would be the total number of treated peers.

Under the exposure mapping framework, a number of simplifications follow. First, the potential outcome collapses from $R_i(z_i, \mathbf{z}_{(-i)})$ to $R_{i}(z_i, g_i)$ where the arguments in the potential outcome do not depend on the sample size $n$. Second, so long as unit $i$'s covariates $\mathbf{X}_i$ are sufficient to control for all confounding, the general interference equivalent of the propensity score in RR is
\begin{equation} \label{eq:PSg}
e(z_{i},g_{i} \mid \mathbf{X}_i ) = P(Z_{i} = z_{i}, G_{i} = g_{i} \mid \mathbf{X}_{i}), \quad{} G_{i} = g(\mathbf{Z}_{(-i)}), \ g:  \{0,1\}^{n-1} \to \reals
\end{equation}
If \eqref{eq:PSg} is known, Proposition 4 of \citet{forastiere2021identification} showed that various causal estimands under general interference can be estimated. Third, there are methods to estimate the density function $e(z_i, g_i \mid \mathbf{X}_{i})$ under some assumptions on the correlation structure between study units; see  \citet{forastiereEstimatingCausalEffects2018} for one example using a Bayesian approach. 

A major drawback of the exposure mapping framework is that investigators need to correctly specify the exposure mapping $g$, a rarity in practice. Nevertheless, the exposure mapping framework has been used in past works to simplify downstream estimation and testing of causal effects. More broadly, a result in Section 4.2 of \citet{HH2008} suggests that a dimension-reducing summary of the peers' treatment vector $\mathbf{Z}_{(-i)}$ may be theoretically necessary to obtain reasonable estimators of standard errors of the estimated causal quantities under partial interference.

Finally, we briefly mention that there is recent work by \citet{li_randomgraphasymptotics_2021}, who proposed another approach to quantify general interference by leveraging advances in statistical network analysis. Specifically, the authors propose to model general interference using an interference graph, a graph that consists of $n$ nodes representing $n$ study units and an $n$ by $n$ symmetric, adjacency matrix $A$ where $A_{ij} = 1$ if unit $i$ and $j$ interfere with each other and $A_{ij} =0$ otherwise; additionally, to avoid self-loops, $A_{ii} = 0$. The adjacency matrix $A$ is assumed to be random and governed by models that are well-studied under statistical network analysis. While the original paper looked at the case without any covariates and the treatment was randomly assigned, we believe incorporating existing models in statistical network analysis is another promising approach to tackle the challenges in propensity score modeling  under general interference.

\section{Final Thoughts}
From theory to application, it's difficult to imagine the field of causal inference without the propensity score and the critical role it plays in studying causal effects of treatment in non-experimental studies. While much of the development on propensity score methods focused on the case where the no-interference assumption holds, many works during the past two decades have tried to relax the no-interference assumption and attempted to extend the propensity score to such settings. While we focused on propensity score modeling in this paper, specifically the challenges that investigators face, we look forward to a future where most, if not all, of these challenges are resolved and the propensity score continues to play a central role in causal inference with and without interference.


\section*{Acknowledgment}

Hyunseung Kang would like to acknowledge support for this project from NSF DMS 1811414.







\bibliographystyle{apa}
\bibliography{sample}

\end{document}